\documentclass{emulateapj}
\usepackage{rotating}
\usepackage{amsmath}
\usepackage{morefloats}
\usepackage{graphicx}
\usepackage{epstopdf}
\usepackage{natbib}
\usepackage{subfigure}
\usepackage{color}
\usepackage[usenames,dvipsnames]{xcolor}

\newcommand{\kext}{\kappa_{\rm ext}}
\newcommand{\zetagal}{\zeta_{\rm gal}}
\newcommand{\sigmak}{\sigma_\kappa}
\newcommand{\kmed}{\kappa_{\rm ext}^{\rm med}}
\newcommand{\zetaq}{\zeta_{q}}

\newcommand{\blens}{B1608$+$656}

\begin{document}

\title{Improving the precision of time-delay cosmography with observations of galaxies along the line of sight}

\shorttitle{Improving time-delay cosmography}
\shortauthors{Greene et al.}

\author{Zach S.~Greene\altaffilmark{1,2},
Sherry H.~Suyu\altaffilmark{1,3,4}, Tommaso Treu\altaffilmark{1,**},
Stefan Hilbert\altaffilmark{3,5}, Matthew W.~Auger\altaffilmark{6},
Thomas E. Collett\altaffilmark{6}, Philip J.~Marshall\altaffilmark{7},
Christopher D.~Fassnacht\altaffilmark{8}, Roger
D. Blandford\altaffilmark{3}, Maru{\v s}a Brada{\v c}
\altaffilmark{8}, L\'eon V.E. Koopmans \altaffilmark{9}}

\email{zg@phys.columbia.edu}
\altaffiltext{1}{Department of Physics, University of California, Santa Barbara, CA 93106-9530, USA}

\altaffiltext{2}{Physics Department, Columbia University, 538 West 120th
Street, 704 Pupin MC 5282, New York, NY 10027, USA}

\altaffiltext{3}{Kavli Institute for Particle Astrophysics and Cosmology, Stanford University, 452 Lomita Mall, Stanford, CA 94035, USA}

\altaffiltext{4}{Institute of Astronomy and Astrophysics, Academia Sinica, P.O.~Box 23-141, Taipei 10617, Taiwan}

\altaffiltext{5}{Max-Planck-Institut f{\"u}r Astrophysik, Karl-Schwarzschild-Str. 1, 85748 Garching, Germany}

\altaffiltext{6}{Institute of Astronomy, University of Cambridge, Madingley Road, Cambridge CB3 0HA, UK}

\altaffiltext{7}{Department of Physics, University of Oxford, Keble Road, Oxford, OX1 3RH, UK}

\altaffiltext{8}{Department of Physics, University of California, One Shields Avenue, Davis, CA, 95616, USA}

\altaffiltext{9}{Kapteyn Astronomical Institute, University of Groningen, PO Box 800 9700 AV Groningen, the Netherlands}

\altaffiltext{**}{Packard Fellow}

\begin{abstract}
In order to use strong gravitational lens time delays to measure
precise and accurate cosmological parameters the effects of mass along
the line of sight must be taken into account.  We present a method to
achieve this by constraining the probability distribution function of
the effective line of sight convergence $\kext$. The method is based
on matching the observed overdensity in the weighted number of
galaxies to that found in mock catalogs with $\kext$ obtained by
ray-tracing through structure formation simulations. We explore
weighting schemes based on projected distance, mass, luminosity, and
redshift.  This additional information reduces the uncertainty of
$\kext$ from $\sigmak\sim$0.06 to $\sim$0.04 for very overdense lines
of sight like that of the system \blens. For more common lines of
sight, $\sigmak$ is reduced to $\lesssim$0.03, corresponding to an
uncertainty of $\lesssim3$\% on distance. This uncertainty has
comparable effects on cosmological parameters to that arising from the
mass model of the deflector and its immediate environment.
Photometric redshifts based on g, r, i and K photometries are
sufficient to constrain $\kext$ almost as well as with spectroscopic
redshifts.  As an illustration, we apply our method to the system
\blens.  Our most reliable $\kext$ estimator gives
$\sigmak$=0.047 down from 0.065 using only galaxy counts.  Although
deeper multi-band observations of the field of
\blens\ are necessary to obtain a more precise estimate, we conclude
that griK photometry, in addition to spectroscopy to characterize the
immediate environment, is an effective way to increase the precision
of time-delay cosmography.
\end{abstract}


\section{INTRODUCTION}
\label{sec:intro}

Strong gravitational lensing can be used to study a number of
important astrophysical topics, ranging from the mass distribution
within galaxies \citep[e.g.,][]{Koc91,T+K02a} and clusters of galaxies
\citep[e.g.,][]{Kne93,Lim++07,San++08} to magnifying distant sources \citep[e.g.,][]{Pet++00,Ell++01,Ric++08}, to cosmography \citep[e.g.,][]{Koc96,Sch++97,Ogu07} \citep[see recent
reviews by][for extensive references]{CSS02,SKW06,Tre10,Bar10,K+N11}.

The power of lensing as a tool for astrophysics stems from its ability
to measure mass independent of its dynamical state, as well as from
its ability to magnify background sources. Detailed mass models of the
deflector galaxies, groups, or clusters of galaxies can be determined
by reproducing the strong lensing observables, i.e. multiple image
time delays, positions, and fluxes \citep[e.g.,][and references
therein]{Kee+10}. However, one of the main limitations of
gravitational lensing is due to the so-called ``mass-sheet
degeneracy'' \citep{FGS85,S+S95,Sah00,Wuc02}.  This emerges from
considering the solution to the lens equation, characterized by a
certain surface mass density $\kappa$ for the deflector in units of
the critical density $\Sigma_{\rm crit}$.  By linear transformations,
one finds that an infinite family of solutions can be obtained. The family
of solutions results in a range of inferred properties both for the
mass model of the main deflector as well as for the properties of the
lensed source.  One needs to break the mass-sheet degeneracy using
physical arguments in order to fully exploit the power of
gravitational lensing.

A number of strategies have been adopted to break this degeneracy both
in the strong and weak lensing regimes
\citep[e.g.,][]{BTP95,Kne++03,BLS04,SBL11}. A common strategy is requiring the
surface mass density of the main deflector to vanish at large
distances from its center. This solution is physically equivalent to
assuming that the distribution of mass along the line of sight (LOS),
excluding the plane of the main deflector, is uniform and equal to the
average of the Universe. This approximation is sufficient for many
applications of gravitational lensing, resulting typically in
uncertainties of only a few percent in the inferred mass distribution
of the main deflector and in the luminosity and size of the lensed
source
\citep{Sel94,KKS97,Koo++06,Tre++09,Hoe++11}.  For higher precision
measurements however, one needs to determine the effects of the line
of sight mass distribution. It is customary to condense these effects
into an equivalent additional mass sheet at the redshift of the main
deflector with uniform surface mass density $\kext$, which can be
positive or negative depending on whether the line of sight is over or
underdense with respect to the average of the Universe
\citep{Sch97}.  In practice, mass-sheet degeneracy can be broken,
i.e., the value of $\kext$ can be inferred, by constraining in an
independent way (1) the mass of the lens galaxy via stellar kinematics
\citep[e.g.,][]{TreuKoopmans02, KoopmansTreu03, BarnabeEtal09,
AugerEtal10, SuyuEtal10, SonnenfeldEtal12}, and/or (2) the total mass
of any intervening mass structures along the line of sight via imaging
and spectroscopy of objects along the line of sight
\citep[e.g.,][]{K+Z04, Fas++06b, Mom++06, SuyuEtal10, WongEtal11,
FKW11}.

Determining $\kext$ is especially important for doing precision
cosmography with gravitational lens time delays. Gravitational lens
time delays are the difference in the arrival time of photons along
the paths corresponding to multiple images arising from geometric and
general relativistic effects. For variable sources, like active
galactic nuclei, time delays can be measured via dedicated monitoring
campaigns \citep[e.g.,][]{Fas++02,Fol++08,ParaficzEtal09, Cou++11,
TewesEtal12}.  If the mass distribution in the plane of the
deflector and along the line of sight is known, the measured time delays can be
converted to the so-called time-delay distance $D_{\Delta t}$, a
combination of three angular diameter distances
\citep[e.g.,][]{Tre10}.  In turn, the time-delay distance contains
information on cosmological parameters, primarily the Hubble constant
$H_0$ \citep{Ref64}, but also the curvature and other parameters
\citep{C+M09,SuyuEtal10,Lin11}. To first approximation, time delays
constrain cosmology in a similar way to baryon acoustic oscillation
experiments and therefore are an excellent complement to other probes
like the cosmic microwave background and type Ia Supernovae
\citep{Lin11,DasLinder12}. It has been shown that with current
techniques and sufficient ancillary data a single gravitational lens
measures the time-delay distance to approximately $5$-$6$\%
\citep{SuyuEtal10, SuyuEtal12}.  

In cases like \blens\ where the time delays are known to $\sim$2\% and
the mass model of the main deflector and its immediate environment is
exquisitely constrained by the data, the dominant source of
uncertainty is the distribution of mass along the line of sight
\cite[e.g.,][]{Bar96}, i.e. effectively the external convergence
$\kext$. Specifically, $D_{\Delta t}\propto(1-\kext)^{-1}$. Thus, for
small $\kext$, the uncertainty $\sigmak$ translates directly into
relative uncertainty in time-delay distance, currently dominating the
overall error budget.

\citet{SuyuEtal10} constrained $\kext$ by using the density of galaxies 
within $45''$ of \blens, which was measured to be twice that of an
average field observed at the same depth by \citet{FKW11}. Then, by
selecting lines of sight with the same galaxy overdensity $\zetagal$
in the Millennium ray-tracing simulations of \citet{HilbertEtal09},
they measured the conditional probability distribution function (PDF),
$P(\kext | \zetagal)$, which was then used as a prior in deriving
cosmological information.  Several efforts are underway to explore
ways to further reduce this source of uncertainty. Those include
spectroscopic and photometric surveys \citep{Mom++06,Wil++08,Aug++07}
as well as measurements of the external shear \citep{SuyuEtal12}

In this paper we focus on refining and developing practical tools to
estimate external convergence by comparing the output of cosmological
numerical simulations with readily available observables such as those
that can be derived from a galaxy photometric catalog.  We define the
relative overdensity $\zeta$ as the value of selected
observables related to the stellar mass, the redshift, and the
projected distance on the sky (angular distance)
divided by the mean of the same observables over all lines of sight.  Extending the
work of~\citet{SuyuEtal10} we consider a set of observables with
relative overdensities $\zeta$ in addition to $\zetagal$, the number
of galaxies along a LOS divided by the average, seeking observables that
minimize $\sigmak$.
In a companion paper, Collett et al. (2013, submitted) a
halo-model approach is used to perform a full reconstruction of the mass
distribution along the line of sight, and thus provide a theoretical
counterpart to the methods and weighting schemes developed here.

The main result of this paper is that using these statistics it is
possible to reduce significantly $\sigmak$, using a multiband
photometric catalog. The amount of gain depends on the specific line
of sight, but for a system like B1608+656 it is possible to reduce it
from 6 to 5\% (even 4\% in the best cases). We note that even reducing
the uncertainty on a single lens from 5 to 4\% is extremely useful
given how rare these strong lenses are and how time-consuming it is to
obtain ancillary data like time delays and high-resolution
images. Oversimplifying for the purpose of illustration, assuming a
target precision of 1\% from a sample of lenses, only 16 systems would
be needed if each were precise to 4\%, as opposed to 25 systems at 5\%
precision each\footnote{In reality, since lenses and sources will have
different redshifts, the likelihoods of the cosmological parameters
from each lens will not be perfectly aligned in multiple dimensions,
and therefore the combined uncertainty should decrease faster than
$1/\sqrt{N}$ \cite[e.g.][]{C+M09,Dob++09,Lin11}.}.

The paper is organized as follows. In Section~\ref{sec:millenium} we
briefly summarize the Millennium Simulation that forms the backbone of
our study. In Section~\ref{sec:counts} we revisit the galaxy count
statistic $\zetagal$. In Section~\ref{sec:weights} we introduce other
statistics involving stellar mass, luminosity, redshift, and
distance. In Section~\ref{sec:observations} we test the influence of
using photometric redshifts instead of spectroscopic redshifts in our
method.  In Section~\ref{sec:fields} we apply our method for several
statistics to the gravitational lens \blens.  Throughout the paper
magnitudes are given on the Vega scale.

\section{SUMMARY OF THE MILLENNIUM SIMULATION}
\label{sec:millenium}

Numerical simulations of large-scale structure provide a way to
determine statistically the amount of external convergence for a lens
system.  By ray tracing through the Millennium Simulation
\citep{SpringelEtal05}, one can produce a simulated image of the sky
populated with galaxies and a corresponding map of the external
convergence, $\kext$, for a given source redshift \citep{HilbertEtal07,
  HilbertEtal08, HilbertEtal09}.  The distribution of $\kext$ values
associated with lines of sight in the simulation that resemble the
lens system of interest (e.g., in terms of the overdensity of galaxies
around the lens system) provides a statistical estimate of the $\kext$
for the lens system.

We use 64 simulated fields of $4\times4$ deg$^2$ from the Millennium
Simulation containing galaxies with positions, redshifts, magnitudes
(in SDSS u,g,r,i,z and 2MASS J,H,K) and stellar masses from the
semianalytic galaxy model of \citet{GuoEtal11}\footnote{Obtained from
the Millennium Database \citep{Lemson++06}.}.  Each field has
approximately 5 million galaxies with redshifts up to $\sim 3.2$ and
an associated map of the external convergence on a $4096\times4096$
grid for a source redshift of 1.4, typical of strong lensing systems
like \blens.  We use each position on the $\kext$ map as a line of
sight, and therefore have approximately $10^9$ lines of sight for the
64 fields. \citet{HilbertEtal09} and \citet{SuyuEtal10} showed that
the distribution of $\kext$ from strong lens lines of sight are very
similar to the distribution for all lines of sight\footnote{The
  $\kext$ constructed from the Millennium Simulation excludes the
  convergence at the primary lens plane (i.e., the redshift of the
  lens galaxy) since this is already accounted for in the lens galaxy
  mass modeling.  Therefore, the external convergence is truly
  external to the lens and is due to line-of-sight contributions.}.  Therefore, we
consider all lines of sight from the 64 fields in our analysis.  These
lines of sight provide the pool from which we select subsets that have
observational properties (based on the galaxies) matching those of the
lens system for determining the $\kext$ of the lens.

\section{GALAXY NUMBER COUNTS AS A PROBE OF $\kext$}
\label{sec:counts}

In this section we revisit the galaxy overdensity statistic introduced
by \citet{SuyuEtal10}.  We recall that the use of relative
overdensities -- instead of absolute densities -- in both the data and
simulations is meant to minimize the impact of theoretical and
observational systematic uncertainties.  In particular this should reduce sensitivity to
the choice of a specific reference simulation~\citep{SuyuEtal10}.

%	\subsection{Galaxy Count Within $45''$ Radius}
%	\label{ssec:n45}

In practice, we study $P(\kext | \zetagal)$, where $\zetagal$ is given
by the number of galaxies within $45''$, $N_{\rm gal}$, in a specific
line of sight, divided by the average value for all lines of sight,
$\overline{N_{\rm gal}}$; i.e.  $\zetagal \equiv \frac{N_{\rm
gal}}{\overline{N_{\rm gal}}}$.  For the Millennium Simulation data,
$\overline{N_{\rm gal}}$ is readily computed by multiplying the total
number of galaxies by $\pi r^{2}/A$ where $r = 45''$ and $A$ is the
angular area of the entire field.  However, in practice one must be
careful of masked regions and edge effects.  The choice of the radius
is motivated by practical reasons. As discussed by \citet{FKW11}, this
is typically the maximum radius that can be surveyed for a target in
middle of one of the two chips, i.e. the standard pointing of the Wide
Field Camera of the Advanced Camera for Surveys (ACS) on board the
\textit{Hubble Space Telescope} (\textit{HST}).  We also impose
restrictions similar to observations of \blens\ such that all galaxies
must have $0 < z < z_{\rm source}$ \citep[where $z_{\rm source}=1.394$
is the source redshift of \blens,][]{Fas++96}, and following
\citet{FKW11}, have $18.5<{\rm mag_{i}}<24.5$.  The results found by
\citet{SuyuEtal10} used only the latter constraint.  In
Section~\ref{sec:fields}, we revisit this field with redshift
information.  Once the PDFs of $\kext$ are computed we define the
width of the PDF $\sigmak$ as the semidifference of the 84 and 16
percentiles of $P(\kext | \zetagal)$.  We remind the reader that in
this paper we use the convergence maps detailed in
Section~\ref{sec:millenium} to supply our $\kext$ values.

The $\kext$ PDF for a LOS with a known relative overdensity in galaxy
count is readily computed by first counting $N_{\rm gal}$ for each
line of sight on the convergence maps.  The convergence maps are
defined on a regular grid with resolution of $\sim$3.5$''$, and the
grid points thus conveniently serve as locations of the lines of
sight for our galaxy number counts.  Then, it is sufficient to 
select lines of sight with overdensity close to the desired value
$\zetagal$.  In practice, we select all lines of sight with
\begin{equation}
\label{eq:ncon}
	\left\lvert N_{\rm gal} - \zetagal \overline{N_{\rm gal}} \right\rvert < E
\end{equation}
where $E$, the interval width, is some integer value.  As we increase $E$, the number of
lines of sight satisfying Equation (\ref{eq:ncon}) also increases.  
As a general rule the parameter $E$
should be chosen to be as small as possible while still leaving
sufficient lines of sight to generate the PDF. Note that sample
variance noise is already introduced by the simulations so we expect
that varying $E$ while keeping it smaller than $\sim \sqrt{N_{\rm
gal}}$ should not affect our results.

\begin{figure}
\centering
\includegraphics[width = 0.5\textwidth]{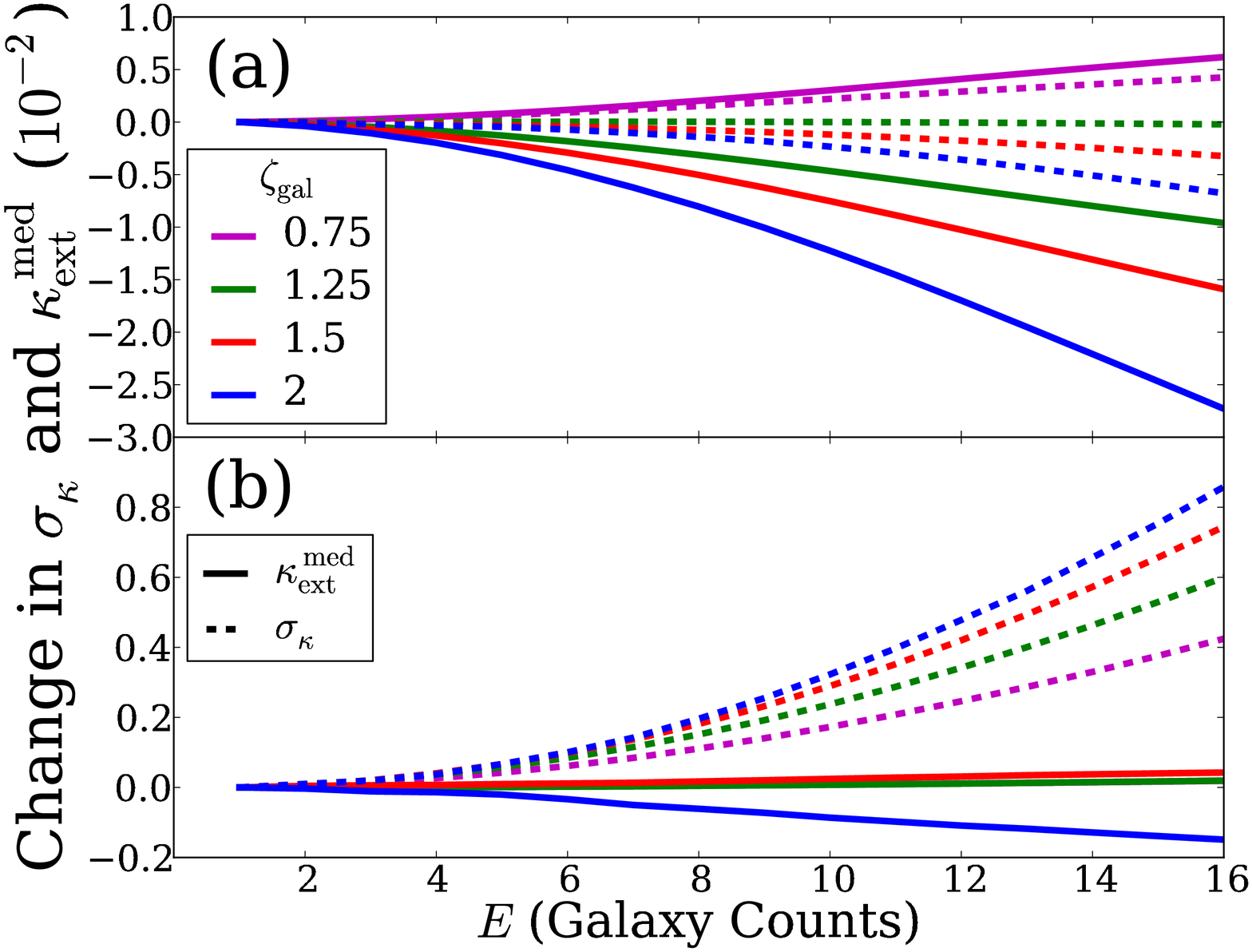}
\caption{\label{fig:echarts} Change in $\kmed$ (solid) and $\sigmak$ (dashed) with $E$ for
$\zetagal = 0.75, 1.25, 1.5$, and $2$ as measured from numerical
simulations.  The change is computed for both (a) unweighted and (b)
weighted PDF combination, by subtracting $\kmed$ and $\sigmak$ for the
given distribution at $E$ by those at $E = 1$.  Unweighted refers to
simply averaging $\kext$ values for all LOS that satisfy
(\ref{eq:ncon}), whereas weighted indicates weighting inversely by the
number of LOS of a particular galaxy count.  The $\kmed$ line for
$\zetagal = 0.75$ in panel (b) follows a nearly identical line to that
of 1.25 and therefore is barely visible.}
\end{figure}

A number of subtleties must be taken into account when constructing
the $\kext$ PDFs. As expected, the distributions are in general highly
skewed. For example, fewer lines of sight become available at higher
relative overdensities (and similarly at lower underdensities; for
conciseness we shall discuss explicitly only the overdensities in our
examples).  Because the number of lines of sight $N_{\rm LOS}$ at a
given galaxy count $N_{\rm gal}$ are always greater than that at
$N_{\rm gal} + 1$, simply constructing a PDF from all $\kext$ whose
corresponding $N_{\rm gal}$ satisfy Equation (\ref{eq:ncon}) would be
biased towards lower $N_{\rm gal}$, and their respective $\kext$
values.  We define $\kmed$ as the median $\kext$ for a given PDF,
and find that a good indicator of the bias in a sample is the change in
$\kmed$ as function of the interval width $E$. The narrowest interval
$E=1$ gives the closest estimate to the true median.  Since lower
$N_{\rm gal}$ pull the entire PDF away from that with a galaxy count
of $\zetagal \overline{N_{\rm gal}}$, we expect that a notable change
in $\kmed$ will occur when we increase $E$.  This is shown in the top panel
of Figure \ref{fig:echarts}.  To circumvent this shift in $\kmed$ that
would lead to a bias in the time-delay distance determination, we
weight each value of $\kext$ by $\frac{1}{N_{\rm LOS}}$ for its
respective $N_{\rm gal}$ -- that is, we find $N_{\rm LOS}$ for the
$N_{\rm gal}$ responsible for contributing a particular $\kext$, and
multiply by the inverse.  Therefore each of the $2E$ $\kext$ PDFs for
a given $N_{\rm gal}$ within $\zetagal
\overline{N_{\rm gal}} \pm E$ carries equal weight, and $\kext$ from
higher $N_{\rm gal}$ offset those from lower, thus ensuring that our
overall distribution remains relatively static.  This is indeed the
case, as the bottom panel of Figure \ref{fig:echarts} shows $\kmed$
declines much more slowly as a function of uncertainty than in the
upper panel.  

However, maintaining a steady median has resulted in increasing the
width of the distribution $\sigmak$. This is expected because PDFs on
either side of $\zetagal \overline{N_{\rm gal}}$ are centered slightly
above or below our target.  For this reason -- and that of residual
effects on the median -- we see that lower galaxy interval widths ($E$)
offer the best results, assuming they can encompass sufficient data to
adequately reduce statistical uncertainty.  For the purposes of this
paper unless otherwise noted we set $E = 2$.

\begin{figure}
	\includegraphics[trim = 0cm 0cm 1cm 1cm, clip = true, width = 0.5\textwidth]{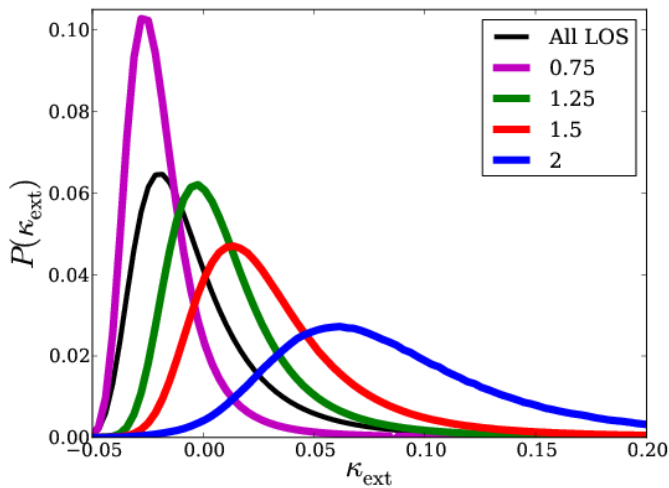}
	\caption{\label{fig:kappapdf} PDFs for $\kext$ constructed from lines of sight that satisfy Equation (\ref{eq:ncon})
	for $\zetagal = 0.75, 1.25, 1.5$, and $2$ and $E = 2$ as measured from numerical simulations.}
\end{figure}

As an illustration of our method, in this paper we investigate the
underdense case $\zetagal = 0.75$ and the overdense cases $\zetagal =
1.25, 1.5$, and $2$ (used in Figure \ref{fig:echarts}), although our
method carries over to any arbitrary value of $\zetagal$.  We show in
Figure~\ref{fig:kappapdf} the PDF of $\kext$ for these
$\zetagal$ values.  As expected, larger $\zetagal$
produces a shift in the PDF along positive $\kext$, however, it also
increases $\sigmak$. In other words more overdense lines of sight have
higher convergence but also a broader range of possible convergences.
In the study of \citet{FKW11}, \blens\ possessed a relative
overdensity of 2.18 without redshift cuts.  We will thus look at the case of
$\zetagal = 2$ with particular interest.

\section{ALTERNATE CHARACTERISTICS AS WEIGHTS}
\label{sec:weights}

Defining $\zetagal = N_{\rm gal}/\overline{N_{\rm gal}}$ is useful for
constructing PDFs of $\kext$ for strong lenses.  Counting $N_{\rm
gal}$ for any line of sight is straightforward and requires only that
each object within $45''$ have $z < z_{\rm source}$ and a flux greater
than an observational limit.  However, by using $N_{\rm gal}$ we
neglect characteristics of an object that may play a significant role
in gravitational lensing (i.e.  mass, redshift, angular offset).  By
using quantities that are closely related to lensing we expect that we
should be able to reduce the uncertainty on $\kext$.  For
example, we do not suspect all LOS with $\zetagal = 2$ to have exactly
the same physical characteristics as \blens, hence we can use relative
overdensities in observable features besides $N_{\rm gal}$ to
construct an even tighter - and perhaps more relevant - PDF.

	\subsection{Redefining $\zeta$ for New Characteristics}
	\label{ssec:ChangingZeta}

To identify LOS with overdensities with particular features, we need
to design a weighting scheme such that all objects are not equivalent, but
weighted by a feature relevant to lensing.  By summing each object
multiplied by its weight within $45''$ we define the weighted sum
\begin{equation}
	\label{eq:weights}
	W_{q} = \sum_{i = 1}^{N_{\rm gal}} q_{i}
\end{equation}
where $q_{i}$ is the weight for object $i$.  Note that for our
definition of $N_{\rm gal}$, $q_{i} = 1$ for all $i$ and therefore the
weighted number of galaxies $W_{\rm gal}$ simply corresponds to $N_{\rm gal}$.  We next
define $\zeta_{q} = {W_{q}}/{\overline{W_{q}}}$ as before.  In
accordance with $\zetagal$, we will set $\zetaq = 0.75, 1.25, 1.5$,
and $2$ for the purpose of demonstrating our general method.

\begin{figure*}[t]
	\includegraphics[width = \textwidth]{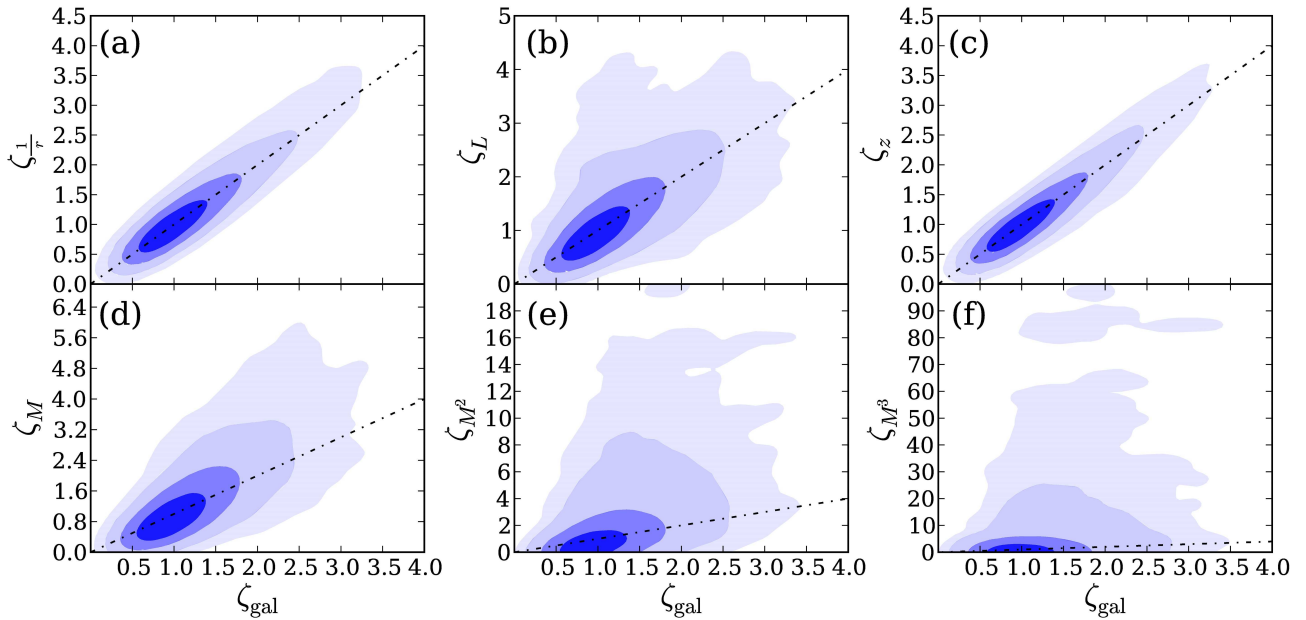}
	\caption{\label{fig:contours} Contour plots showing $\zetagal$
	vs. $\zetaq$ for various weighting methods, as measured from numerical simulations.  Dark to lighter
	shades refer to $1, 2, 3$, and $4 \sigma$ regions,
	respectively.  The dotted line marks $\zetagal = \zetaq$.  We
	see that the different $\zetaq$ are correlated,
	i.e. under/overdense LOSs in one metric are typically
	under/overdense in other metrics as well. However, they are
	not identical, implying that imposing multiple $\zetaq$
	conditions adds information and thus sharpens the PDF.}
\end{figure*}

		\subsubsection{Weighting by Radius}
		\label{sssec:radius}

The angular separation between the source and a nearby object is a significant
factor in distorting, and thereby shaping, the path through which the source's
light passes.  We expect then that each galaxy within $45''$ does not
contribute equally, but that those nearer the optical axis of the lens
are more influential than those farther away. In particular for
isothermal total mass distributions \citep[appropriate for massive
galaxies or around the scale radius of halos,
e.g.,][]{Gav++07,Lag++10}, we expect the convergence to decline as the
inverse projected distance from the deflector. This gives rise to our
first weighting method of $1/r$.  We will scale all objects that
satisfy $10'' < r \leq 45''$ by $1/r$.  At $r \leq 10''$ weighting
becomes sensitive to small changes in $r$.  Thus for $r \leq 10''$ we
allow each object to carry a weight of $1/10$, giving us a continuous
weighting function.  We note that in general objects that are very
close to the main deflector are more likely to be physically
associated with it and exert a stronger impact
\cite[e.g.,][]{K+Z04}. Therefore it is prudent to model them
explicitly, rather than considering them as part of the statistical
lines of sight effect. This might require obtaining as much
information as possible on them, including redshifts
\citep{Mom++06,Aug++07} and possibly stellar velocity dispersions,
especially if they are consistent with massive galaxies.

		\subsubsection{Weighting by Redshift}
		\label{sssec:redshift}

Objects close to the source along the line of sight have minimal
lensing effects from the scaled deflection of light rays.  Likewise,
those nearest to the observer are relatively insignificant.  In order
to approximately account for this, our next heuristic weighting method
is quadratic in $z$ and defined as $z_{\rm source}\cdot z - z^{2}$ where
$z_{\rm source}$ and $z$ are the redshifts of the source and the
object along the LOS, respectively.  For simplicity the notation of
this weighting method is ``$z$''.

		\subsubsection{Weighting by Stellar Mass}
		\label{sssec:mass}

The most massive galaxies will produce detectable lensing effects over
a larger area of the sky.  Thus we choose one of the weighting methods
to be $M^{n}$ where $M$ is the object's stellar mass and $n$ is some
positive integer.  The rationale for this scaling is that at the high
mass end of the galaxy stellar mass function, the relation between
stellar mass and total mass is non-linear. According to, e.g., weak
lensing, clustering, satellites, and abundance matching studies, the
total mass increases with stellar mass faster than linear in for the
most massive galaxies \citep{Man++06,Wak++11,Beh++10,Lea++12,Mor++11}.
This is consistent with the fact that the central galaxies of massive
clusters and poor groups do not typically differ in stellar mass by
orders of magnitude even though their halos do.  For $M^{n}$ with $n >
1$ we will consider both $\sum_{i = 1}^{N_{\rm gal}} M_{i}^{n}$ and
$\sqrt[n]{\sum_{i = 1}^{N_{\rm gal}} M_{i}^{n}}$.  The former will be
denoted as $W_{M^{n}}$ while the latter as the root sum of the squares
$W_{M_{\rm rss}^{n}}$.  In this paper we explore the cases of $n = 1,
2,$ and $3$.  In this section we assume to know precisely the correct
masses, as given by the Millennium catalog; in
Section~\ref{sec:observations} we allow each mass to depend on its
respective photometric redshift to assess the reliability of a $M^{n}$
weighting method.

		\subsubsection{Weighting by Luminosity}
		\label{sssec:luminosity}

In practice, inferring the stellar mass of an object with limited
observational data can be difficult, leading to large uncertainties.
For this purpose we will also explore weighting by luminosity, $W_{L}$.

	\subsection{$\kext$ PDF with New Statistics}
	\label{ssec:stats}

We now proceed with the method outlined in Section~\ref{sec:counts}, requiring that
$\zetaq = 0.75, 1.25, 1.5$, or $2$
for all aforementioned $q$.  Contour plots for $\zetaq$ versus $\zetagal$
given in Figure \ref{fig:contours} show how the various relative overdensities
are related.  Because each weighting scheme differs in $\overline{W_{q}}$
from every other, we would like to ensure that keeping
a consistent interval width $E$ does not affect our relative spread.  Weighting
schemes with low $\overline{W_{q}}$ would offer a higher percentage of total
lines of sight than those with large $\overline{W_{q}}$.  To normalize our spread we
multiply each $W_{q}$ by
$\frac{\overline{N_{\rm gal}}}{\overline{W_{q}}}$.  Thus, we
generalize Equation
(\ref{eq:ncon}) to the following for new statistics:
\begin{equation}
\label{eq:wcon}
	\left\lvert \frac{\overline{N_{\rm gal}}}{\overline{W_{q}}} W_{q} - \zetaq \overline{N_{\rm gal}} \right\rvert < E.
\end{equation}
Furthermore, $W_{q}$ is no longer restricted to discrete integer
values, but rather a continuum.  Still, we expect $\frac{d N_{\rm
LOS}}{d W_{q}} \neq 0$ at under- and overdense $\zetaq$ so it is
necessary to normalize by the inverse number of LOS.  We allow $2E$
bins, each of length 1, from $\zeta_{q} \overline{N_{\rm gal}} - E$
to $\zeta_{q} \overline{N_{\rm gal}} +E$ (as previously done in Section~\ref{sec:counts} with
discretely-valued $N_{\rm gal}$), and define $N_{\rm LOS}$ as the number
of LOS within a particular bin.  We then weight each $\kext$ value
by $\frac{1}{N_{\rm LOS}}$ of its respective bin when constructing
the PDF.

\begin{figure*}
	\includegraphics[width = \textwidth]{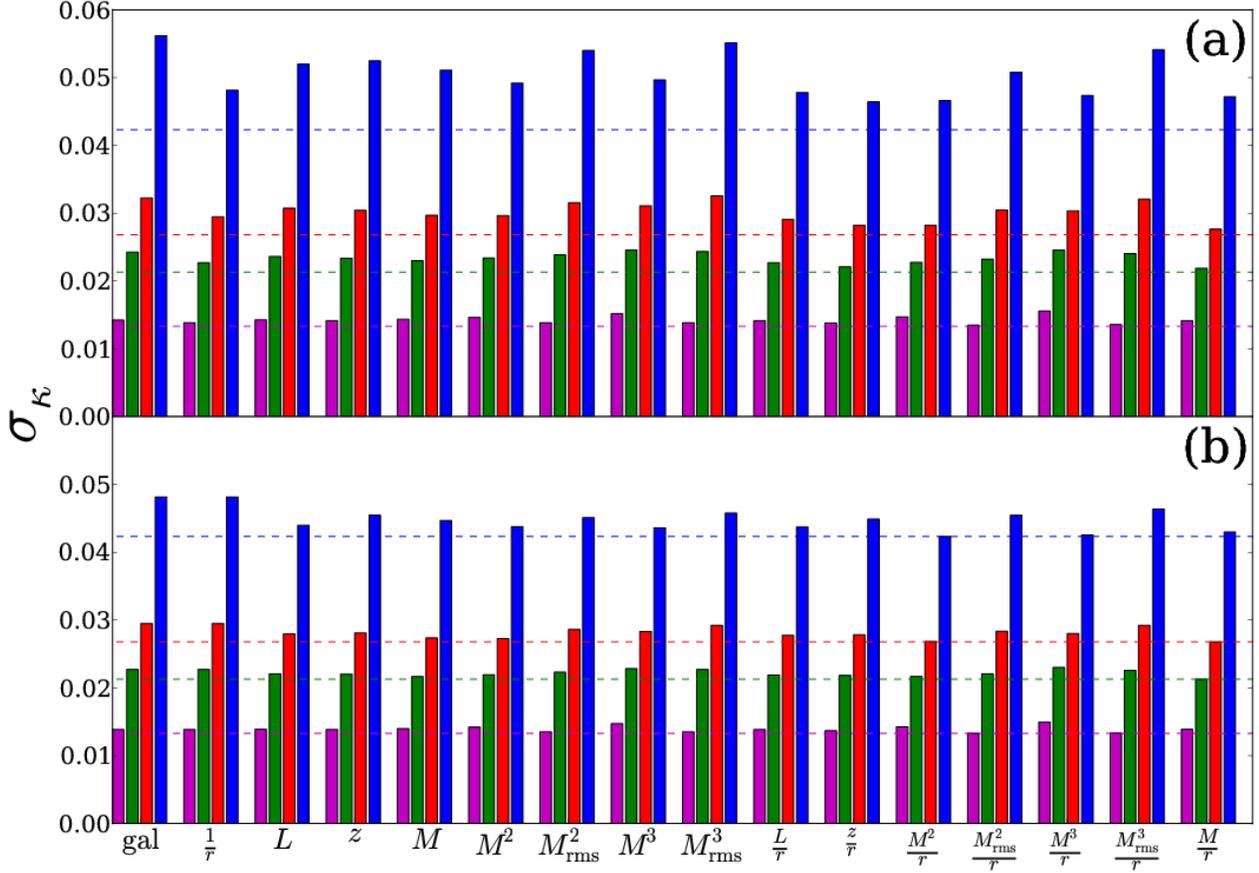}
	\caption{\label{fig:weightsbar} Values for $\sigmak$ for various weighting schemes
	with (a) successive $\zetagal$ and $\zetaq$ conditions, and
        (b) successive $\zetagal$, $\zetaq$ and $\zeta_{\frac{1}{r}}$
	conditions as measured from numerical simulations.  As discussed in
	the text, note that cases where multiple conditions become equivalent
	(such as the leftmost bar in the top panel, corresponding to imposing
	two consecutive $\zetagal$ conditions) amount to a redundancy and no
	new information is gained.  The color scheme is the same as in previous
        plots: purple, green, red, and blue refer to $\zetaq = 0.75, 1.25, 1.5,$
	and $2$, respectively.  The horizontal dashed lines correspond to the lowest value of $\sigmak$ obtained through the weighting schemes considered.  We find the uncertainty on $\kext$ is reduced by
	imposing more conditions.}
\end{figure*}

The variables discussed in Section~\ref{ssec:ChangingZeta} are the principal contributors to
gravitational lensing; however, using a single variable may be too basic of an
approximation.  We can expand our definition of Equation (\ref{eq:weights}) to
allow the weighted sum ($W_{q}$) to be the product of characteristics
\begin{equation}
\label{eq:multiweights}
	W_{q} = \sum_{i = 1}^{N_{\rm gal}} \prod_{j = 1}^{n_{\rm var}} q_{ij}
\end{equation}
where $n_{\rm var}$ is the number of variables.  Figure
\ref{fig:weightsbar} (the seven right-most set of bars) and Table
\ref{wsigs} show that this leads to greater improvements in $\sigmak$
when combining $1/r$ with the established $q$, though in principle
this can be done with any combination.

In addition to generalizing $W_{q}$, we can use Equation
(\ref{eq:wcon}) to impose multiple $\zetaq$ conditions.  For
example, we can require that LOS satisfy Equation (\ref{eq:wcon}) for
both galaxy count and $1/r$ weighting.  This is a more stringent
demand, as LOS must now pass two separate cuts.  Note that our
respective $\zetaq$ (in this example $\zetagal$,
$\zeta_{\frac{1}{r}}$) values need not be equivalent. However, for the sake of illustration in this paper we will assume they
are.  We refer to the number of applied conditions as $n_{\rm con}$.
This imposes consecutive cuts that improve the quality of the $\kext$
PDF, but reduce the number of LOS.  As long as $n_{\rm con}$ is not
chosen to be large enough to introduce statistical uncertainty, we
expect a sharpening of the peak as remaining LOS will be more relevant
to the lens of interest.  We must, however, now expand our definition
of $N_{\rm LOS}$ to incorporate combinations of every possible value
between $\pm E$ for each of $n_{\rm con}$.  Thus each LOS does not
simply correspond to one of $2E$ values, but instead
$\left(2E\right)^{n_{\rm con}}$.

We remind the reader that $\zetaq$ refers to the
relative overdensity of any of the aforementioned weights (e.g. galaxy counts,
$1/r$, $z$, $L$, $M$).  Using the prescribed method we measure $\sigmak$
for each weighting scheme.  We do this by constructing the PDF from LOS that satisfy
Equation (\ref{eq:wcon}) in three variations: (1) $\zetaq$; (2) $\zetaq$ and $\zetagal$,
and; (3) $\zetaq, \zetagal$, and $\zeta_{\frac{1}{r}}$.  It is worth noting that
for case (2) this amounts to applying the same condition twice for $\zetagal$,
and in case (3) we have a similar redundancy for both $\zetagal$ and $\zeta{\frac{1}{r}}$.

We expect that imposing more conditions, as in cases (2) and (3),
would lead to a smaller width of the $\kext$ PDF.
Figure~\ref{fig:weightsbar} shows $\sigmak$ for these latter two
cases.  There are several features here worth noting.  First, changes
in $\sigmak$ for $\zetaq = 0.75, 1.25$, and $1.5$ are relatively small
when compared with those for $\zetaq = 2$. Thus, it is most easy to
detect any increase or decline in $\sigmak$ for $\zetaq = 2$.  Second,
if we assume that any change in $\sigmak$ for each $q$ at $\zetaq = 2$
is indicative of the change at lower $\zetaq$ (albeit on a smaller
scale), then we can restrict the analysis to $\zetaq = 2$ to determine
which variables constrain $\kext$ the most.

We also see that in all cases $\sigmak$ decreases when more conditions are
imposed, as expected.  Table~\ref{wsigs} gives the values for the bars
in Figure~\ref{fig:weightsbar}, along with $\zetaq$ for case (1) as
mentioned above.  We note that for lenses with $\zetaq\lesssim1.5$ in
multiple conditions, the uncertainty in $\kext$ is reduced to
$\lesssim0.03$, a level that is comparable to or smaller than the
strong lens mass modeling uncertainty in terms of its impact on the
time-delay distance \citep[e.g.][]{SuyuEtal10, SuyuEtal12}.  As
current and future surveys are expected to discover at least hundreds
of lenses \citep{OguriMarshall10}, we expect an efficient sample
for cosmographic studies to contain lenses with relative
overdensities $\lesssim 1.5$.  This will ensure that uncertainties due
to the LOS structures are subdominant.

\setlength{\tabcolsep}{0.15cm}

\begin{table*}[t]
	{\footnotesize
	\begin{center}
	\caption{\label{wsigs} Values of $\sigmak$ for $\zetaq = 2$ for various
	weighting methods and conditions.}
	\begin{tabular}{cccccc}
		\hline
		$q$ & $\zetaq = 2$ & $\zetagal,\zetaq = 2$ & $\zetagal,\zeta_{\frac{q}{r}} = 2$ & $\zetagal,\zetaq,\zeta_{\frac{1}{r}} = 2$ & $\zetagal,\zeta_{\frac{q}{r}},\zeta_{\frac{1}{r}} = 2$\\
		\hline
		$\rm gal$ & 0.0562$\pm$0.0003 & | & | & | & |\\
		$\frac{1}{r}$ & 0.0534$\pm$0.0002 & 0.0481$\pm$0.0003 & | & | & |\\
		$L$ & 0.0464$\pm$0.0002 & 0.0520$\pm$0.0004 & 0.0478$\pm$0.0003 & 0.0440$\pm$0.0005 & 0.0437$\pm$0.0004\\
		$z$ & 0.0507$\pm$0.0003 & 0.0525$\pm$0.0004 & 0.0464$\pm$0.0004 & 0.0455$\pm$0.0004 & 0.0449$\pm$0.0003\\
		$M$ & 0.0382$\pm$0.0002 & 0.0511$\pm$0.0006 & 0.0472$\pm$0.0004 & 0.0447$\pm$0.0006 & 0.0430$\pm$0.0005\\
		$M^{2}$ & 0.0319$\pm$0.0001 & 0.0492$\pm$0.0006 & 0.0466$\pm$0.0006 & 0.0438$\pm$0.0009 & 0.0423$\pm$0.0006\\
		$M_{\rm rss}^{2}$ & 0.0449$\pm$0.0002 & 0.0540$\pm$0.0005 & 0.0508$\pm$0.0004 & 0.0451$\pm$0.0007 & 0.0454$\pm$0.0005\\
		$M^{3}$ & 0.0319$\pm$0.0001 & 0.0496$\pm$0.0008 & 0.0474$\pm$0.0005 & 0.0436$\pm$0.0011 & 0.0425$\pm$0.0008\\
		$M_{\rm rss}^{3}$ & 0.0417$\pm$0.0002 & 0.0551$\pm$0.0007 & 0.0541$\pm$0.0005 & 0.0458$\pm$0.0008 & 0.0464$\pm$0.0005\\
		\hline
	\end{tabular}
      \end{center}
	Notes. The $\sigmak$ measured from numerical simulations decrease with some of the unique condition
	$\zetaq = 2$ (second column) because they incorporate a large range of $\zetagal$ values not
	necessarily correspondent to $\zetagal = 2$.  This problem is fixed by always imposing
	initial condition $\zetagal = 2$ (columns
        3--6).} 
\end{table*}

The spread of $\kext$ for $\zetaq = 2$ in Table \ref{wsigs} is easily
explained by Figure \ref{fig:contours}.  If we look at $\zetaq = 2$
for $\zetaq \neq \zetagal$ we see that a majority of its $\zetagal$
values lie at $\zetagal < 2$.  This causes a shift in the overall
$\kext$ distribution, lowering the median and shrinking $\sigmak$.
Therefore, such a decrease is not the result of improving our method to
find $\kext$, but the inclusion of a large number of small $\kext$
values with $\zetagal < 2$.  This reiterates the effectiveness of
imposing $\zetagal$ and other $\zetaq$ conditions, as in Figure
\ref{fig:weightsbar} and the remaining columns of Table \ref{wsigs}.  

While all weighting methods lead to a decrease in $\sigmak$, the
lowest values are for $W_{\frac{M^{n}}{r}}$.  For the $\kext$ PDFs that
include the $\zetagal$ constraint, the weighting scheme that leads to
the tightest PDF is $\zetagal, \zeta_{\frac{M^{2}}{r}},
\zeta_{\frac{1}{r}} = 2$ with the corresponding PDF width as $\sigmak 
= 0.0423 \pm 0.0006$ (see columns 3 to 6 in Table \ref{wsigs}).  This is
a substantial drop from our initial finding of $\sigmak = 0.0562 \pm
0.0003$ for $\zetagal = 2$, and $\sigmak\sim0.065$ that was obtained
by \citet{SuyuEtal10} for B1608 without applying the $z<1.4$ cut.

\section{Testing the fidelity of $\kext$ estimates based on photometric redshifts}
\label{sec:observations}

Until this point our simulated galaxy catalogs from the Millennium
Simulation have allowed us to neglect uncertainties in redshift and
stellar mass that would normally arise from observations.  However,
getting spectroscopic redshifts for a large sample of objects -- many
at $z > 1$ -- is difficult and expensive.  It is thus prudent to focus
spectroscopic resources on the brighter objects and those closer to
the main deflector, while using photometric redshifts for the
remaining objects along the line of sight. Because a galaxy's
estimated stellar mass and luminosity are dependent on its redshift,
their uncertainties are sensitive to errors in $z$.  It is necessary
then to estimate the uncertainties in $\kext$ associated with
obtaining redshifts photometrically.

\subsection{Estimating photometric redshifts}
\label{ssec:bpz}

The Millennium Simulation gives magnitudes for SDSS u, g, r, i, z and
2MASS J, H, K bands.  We use the Bayesian Photometric Redshift
\citep[hereafter BPZ;][]{2000ApJ...536..571B,Coe++06} program to 
calculate the photometric redshift $z_{\rm phot}$, which is defined as
the peak of the redshift PDF, for all objects.  To evaluate how
$\kext$ is affected by the quality of $z_{\rm phot}$, we examine three
different band combinations.  First of all, we use ugrizJHK to compute
what we may assume to be the best approximation to $z_{\rm spec}$.
Secondly, we use g, r, i, and K bands, in an effort to strike a
compromise between survey speed and wavelength coverage, as a measure
of the effectiveness of our method when only a few bands are
available.  Lastly, we compute $z_{\rm phot}$ using just g, r, and i
bands to evaluate how our method holds under the least number of bands
from which a redshift might be computed.  Typically, optical bands
such as g, r, and i are the most readily available or do not require
long integration times, which make them ideal for large-scale surveys.

	\subsection{Calculating $\kmed$ based on galaxy number counts with photometric data}
	\label{ssec:allbands}

The errors associated with photometric redshifts are expected to
decrease as wavelength coverage is increased. However, in practice
obtaining ugrizJHK is observationally expensive.  Thus this section
serves primarily as a premise for the optimal strategies associated
with photometrically determining $z$.

\begin{table*}[t]
        {\footnotesize
	\caption{Values of the shifts $\Delta \kmed$ and $\Delta \sigmak$ with $\zetaq = 2$ for $z_{\rm phot}$ computed using a variety of band combinations.}
	\label{zphotsigs}
	\begin{center}
	\begin{tabular}{ccccccc}
		\hline
		$q$ & \multicolumn{2}{c}{$\rm ugrizJHK$} & \multicolumn{2}{c}{$\rm griK$} & \multicolumn{2}{c}{$\rm gri$}\\
		\cline{2-7}
		& $\Delta \kmed$ & $\Delta \sigmak$ & $\Delta \kmed$ & $\Delta \sigmak$ & $\Delta \kmed$ & $\Delta \sigmak$\\
		\hline
		$\rm gal$ & 0.0060$\pm$0.0003 & 0.0023$\pm$0.0002 & 0.0066$\pm$0.0005 & 0.0026$\pm$0.0005 & 0.0155$\pm$0.0008 & 0.0063$\pm$0.0005\\
		$\frac{1}{r}$ & 0.0075$\pm$0.0003 & 0.0034$\pm$0.0003 & 0.0079$\pm$0.0005 & 0.0025$\pm$0.0005 & 0.0151$\pm$0.0008 & 0.0051$\pm$0.0005\\
		$L$ & 0.0071$\pm$0.0008 & 0.0041$\pm$0.0009 & 0.0075$\pm$0.0011 & 0.0038$\pm$0.0019 & 0.0145$\pm$0.0013 & 0.0060$\pm$0.0019\\
		$z$ & 0.0063$\pm$0.0011 & 0.0039$\pm$0.0007 & 0.0127$\pm$0.0017 & 0.0086$\pm$0.0013 & 0.0219$\pm$0.0021 & 0.0116$\pm$0.0013\\
		$M$ & 0.0054$\pm$0.0012 & 0.0022$\pm$0.0011 & 0.0025$\pm$0.0009 & 0.0020$\pm$0.0015 & 0.0168$\pm$0.0011 & 0.0065$\pm$0.0015\\
		$M^{2}$ & 0.0067$\pm$0.0022 & 0.0029$\pm$0.0028 & 0.0050$\pm$0.0008 & 0.0028$\pm$0.0022 & 0.0142$\pm$0.0008 & 0.0042$\pm$0.0022\\
		$M_{\rm rss}^{2}$ & 0.0103$\pm$0.0008 & 0.0047$\pm$0.0021 & 0.0105$\pm$0.0027 & 0.0052$\pm$0.0023 & 0.0202$\pm$0.0025 & 0.0076$\pm$0.0023\\
		$M^{3}$ & 0.0047$\pm$0.0015 & 0.0041$\pm$0.0029 & 0.0159$\pm$0.0059 & 0.0085$\pm$0.0021 & 0.0148$\pm$0.0025 & 0.0068$\pm$0.0021\\
		$M_{\rm rss}^{3}$ & 0.0080$\pm$0.0012 & 0.0064$\pm$0.0015 & 0.0124$\pm$0.0046 & 0.0033$\pm$0.0019 & 0.0190$\pm$0.0012 & 0.0085$\pm$0.0019\\
		$\frac{L}{r}$ & 0.0077$\pm$0.0012 & 0.0041$\pm$0.0007 & 0.0077$\pm$0.0013 & 0.0029$\pm$0.0016 & 0.0115$\pm$0.0015 & 0.0032$\pm$0.0016\\
		$\frac{z}{r}$ & 0.0025$\pm$0.0007 & 0.0017$\pm$0.0009 & 0.0047$\pm$0.0008 & 0.0026$\pm$0.0007 & 0.0143$\pm$0.0011 & 0.0045$\pm$0.0007\\
		$\frac{M^{2}}{r}$ & 0.0061$\pm$0.0009 & 0.0028$\pm$0.0007 & 0.0077$\pm$0.0011 & 0.0021$\pm$0.0015 & 0.0122$\pm$0.0015 & 0.0043$\pm$0.0015\\
		$\frac{M_{\rm rss}^{2}}{r}$ & 0.0116$\pm$0.0010 & 0.0049$\pm$0.0007 & 0.0121$\pm$0.0017 & 0.0049$\pm$0.0017 & 0.0203$\pm$0.0017 & 0.0072$\pm$0.0017\\
		$\frac{M^{3}}{r}$ & 0.0085$\pm$0.0007 & 0.0047$\pm$0.0009 & 0.0264$\pm$0.0037 & 0.0131$\pm$0.0017 & 0.0128$\pm$0.0018 & 0.0036$\pm$0.0017\\
		$\frac{M_{\rm rss}^{3}}{r}$ & 0.0109$\pm$0.0013 & 0.0053$\pm$0.0017 & 0.0100$\pm$0.0019 & 0.0071$\pm$0.0021 & 0.0136$\pm$0.0023 & 0.0051$\pm$0.0021\\
		$\frac{M}{r}$ & 0.0056$\pm$0.0006 & 0.0015$\pm$0.0009 & 0.0017$\pm$0.0011 & 0.0006$\pm$0.0017 & 0.0194$\pm$0.0012 & 0.0051$\pm$0.0017\\
		\hline
	\end{tabular}
      \end{center}}
\end{table*}

\begin{figure}[b]
	\includegraphics[width = 0.5\textwidth]{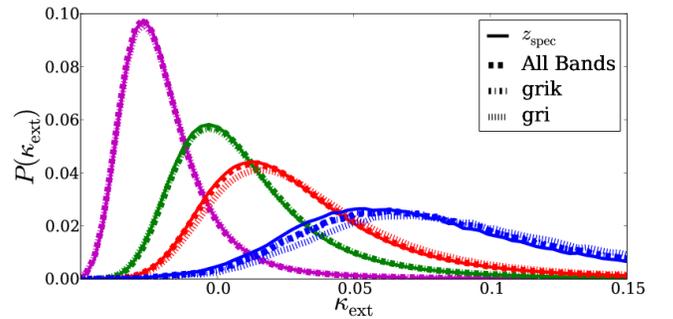}
	\caption{\label{fig:zphot} $\kext$ PDFs constructed from LOS
	satisfying Equation (\ref{eq:ncon}) for $\zetagal = 0.75$ (purple),
	$1.25$ (green), $1.5$ (red), and $2$ (blue) for spectroscopic (solid) and photometric
	redshifts as measured from numerical simulations.  Photometric redshifts are computed using all
	available optical and infrared bands (dashed), griK bands
	(dot-dashed; shown for $\zetagal=2$ only) and gri bands
	(dotted).}
\end{figure}

In Figure~\ref{fig:zphot}, we plot $\kext$ PDFs for $N_{\rm gal}$ with
the original requirement of Equation \eqref{eq:ncon} for $\zetagal =
0.75, 1.25, 1.5$, and $2$ based on the spectroscopic redshifts (solid)
and various photometric-redshift estimations (dashed, dot-dashed,
dotted).  With photometric-$z$ computed using all 8-bands, we recover
the overall shape of the $\kext$ PDF.  The $\kext$ PDF based on griK
is shown only for $\zetagal = 2$, which is nearly indistinguishable
from the one based on all bands.  It is evident that fewer photometric
bands causes a shift in $\kext$ toward higher values.  This is
consistent with the fact that with only gri BPZ tends to produce
photo-z with large uncertainties and slightly low bias. Thus, high
redshift objects are incorrectly assigned $z<1.4$ and vice versa, but
the net exchange favors an increase in objects with $z_{\rm phot} <
1.4$.  If we assume these underestimated high-$z$ objects are unlikely
to be correlated with already overdense regions for $z_{\rm spec} <
1.4$ (a reasonable assumption), then a uniform increment, $\delta
N_{\rm gal}$, is accounted for along each LOS.  Thus our LOS of
previous relative overdensities go from $\zetagal \overline{N_{\rm
gal}} \rightarrow \zetagal \overline{N_{\rm gal}} + \delta N_{\rm
gal}$, while our new mean becomes $\overline{N_{\rm gal}^{\prime}} =
\overline{N_{\rm gal}} +
\delta N_{\rm gal}$.  Multiplying $\zetagal$ by
$\overline{N_{\rm gal}^{\prime}}$, we find that our new count based on $z_{\rm phot}$ and satisfying
\eqref{eq:ncon} is
$\zeta_{\rm gal} (\overline{N_{\rm gal}} + \delta N_{\rm gal})$,
having $\delta N_{\rm gal} \left(\zetagal - 1 \right)$ more galaxies
than the previous selection criteria.  Consequently, relative galaxy
(over/under)\-densities based on photometric redshifts are associated
with more (over/under)\-dense lines of sight compared to
spectroscopy-based galaxy overdensities with the same nominal
value. If one were to ignore the difference between photometric and
spectroscopic redshifts, this would induce a bias in the estimation of
$\kext$. This bias is small when using ugrizJHK or even griK, but
notable when only gri is available. It is possible, however, to
further reduce such a bias by improving the photo-z to remove the
small bias or by using exactly the same method of redshift
determination in the actual observations and the simulations.

The accuracy of the $\kext$ PDF should reflect the effectiveness of a
band combination's ability to estimate correctly an object's redshift.  To
quantify this accuracy we define the change in PDF width $\Delta \sigmak \equiv
\sigmak^{\rm phot} - \sigmak^{\rm true}$ where $\sigmak^{\rm phot}$
and $\sigmak^{\rm true}$ are the uncertainty in $\kext$ for $z_{\rm
phot}$ and $z_{\rm spec}$, respectively.  Similarly, we define $\Delta
\kmed$ as the difference between the photometrically-determined
$\kmed$ and the spectroscopically-determined
$\kmed$.  We expect these quantities to be the
smallest for $z_{\rm phot}$ with all 8 bands, and to increase as fewer
bands are used in constructing the redshift.  The first row of Table
\ref{zphotsigs} shows that with all 8 bands or with even only griK,
the change in the $\kext$ PDF is $<$0.007, corresponding to $<$0.7\%
impact on the time-delay distance. {\sl We thus conclude that the minimal
set of filters necessary to achieve a 1\% precision and accuracy on
time-delay distance \citep[see][for a summary of cosmological
implications]{SuyuEtal12a} is three optical filters plus K.}

	\subsection{Impact of photometric redshifts on accuracy and precision of $\kext$ estimates using multiple weights}
	\label{ssec:photweights}

Because Section \ref{sec:weights} demonstrated that using
characteristic features of galaxies provides a sharper PDF we need to
explore how photometric redshifts affect these weights and their
respective PDFs.  Specifically, we would like to confirm our intuition
that these weighted PDFs behave in the same manner as their
spectroscopic counterparts, so that we may choose a universal optimal
weighting method that is independent of how an object's attributes are
obtained.

We use our photometric redshifts for $W_{z}$ and rescale the Millennium
Simulation masses
and luminosities by $\left(\frac{d_{\rm L}\left(z_{\rm
phot}\right)}{d_{\rm L}\left(z_{\rm spec}\right)}\right)^{2}$ where
$d_{\rm L}\left(z\right)$ is the luminosity distance at
redshift $z$.  This is a reasonable approximation for stellar mass, as
to first order it scales proportional to luminosity.
Table \ref{zphotsigs} lists the changes in the median $\Delta \kmed = \kext^{\rm med, phot} - \kext^{\rm med,spec}$ and spread $\Delta \sigmak = \sigmak^{\rm phot} - \sigmak^{\rm spec}$ of the convergence distributions.  We find that for nearly all $q$, $\Delta \sigmak$ increases with fewer bands, which is consistent with our observations for $\zetagal$. However, when using either all bands or griK, the changes are below $0.01$ for a majority of the $q$. 

We thus conclude that, as in the previous section, one should use as
much information as possible to infer $P(\kext)$ for the observed line
of sight. As shown in Figure~\ref{fig:contours}, observables like
position, luminosity, redshift, and stellar mass add valuable
information and can improve both the precision (by reducing $\sigmak$)
and the accuracy (by shifting $\kmed$ closer to the ``true'' value) of
the inference.  In case spectroscopic redshifts are not available,
we recommend using at least three optical bands and one infrared band
for the weighting schemes considered in this paper.
For the current level of cosmological precision and accuracy, using
griK is sufficient to constrain $\kext$ almost as well as
with spectroscopic redshifts.

\section{ILLUSTRATING THE METHOD WITH THE CASE STUDY B1608+656}
\label{sec:fields}

The previous sections outlined a new approach for determining $\kext$
for an arbitrary lens given that sufficient characteristics are known
to calculate the relevant $\zetaq$.  In this section we illustrate how
the method works in practice using \blens\ as a case study. The data
on the \blens\ field includes deep HST imaging in F555W and F814W (9
and 11 orbits, respectively; GO-10158, PI Fassnacht), as well as more
shallow imaging in Gunn g, r, and i obtained with the Palomar 60-Inch
Telescope \citep[full details of the observations can be found
in][]{Fas++06b}.  This section should be taken as an illustration
only, since the data in hand for \blens\ are not sufficient to achieve
the full potential of our method. Therefore we do not revisit the
cosmological implications of \blens\ in this work. Work is in progress
to collect the necessary photometry and spectroscopy and future papers
will present improved estimates of $\kext$ and cosmological parameters
for \blens\ and other systems.

\subsection{Field Preparation and $W_{q}$}
\label{ssec:prep}

As a reference, we use the central portion ($1.1 \times 1.1$
deg$^{2}$) of the COSMOS field \citep[Cosmic Evolution
Survey;][]{ScovilleEtal07} as a sample for measuring the average
number of galaxies and also the average properties of the
features for all lines of sight, i.e., $\overline{W_{q}}$.  
COSMOS data, like \blens\ data, has ACS F814W photometry that is
sufficiently deep to satisfy the upper limit of the $I$ magnitude cut.
We use the 2006 ACS Catalog \citep{2007ApJS..172..219L} and match the
galaxies to those in the 2006 Photometry Catalog
\citep{2007ApJS..172...99C} in order to obtain redshifts and stellar
masses.  We consider two objects in opposing catalogs to be identical
if they have angular separation $\le 0.5''$.  Of the $\sim 124,000$
objects in the ACS databank with $18.5 < m_{\rm F814} < 24.5$,
approximately $118,000$ or $95\%$ have photometric coordinate-matched
counterparts.  Only $\sim 60$ ACS objects are found to be $\le 0.5''$
to two different objects from the photometry catalog. In these few
cases of ambiguous identification, the stellar mass from the multiband
photometric catalog \citep{Ilb++10} is assigned to objects in the ACS
catalog proportional to their fractional contribution to the total
flux.

Because a small subsample of the ACS catalog does not have matches in
the photometric database, we expect our mean relative overdensity
values for all but $\zetagal$ and $\zeta_{1/r}$ to be slightly
underestimated.  As a solution, a correction factor $b = N_{\rm
ACS}/N_{\rm phot}$ is applied where $N_{\rm ACS}$ and $N_{\rm phot}$
are the total and matched number of galaxies in the ACS catalog,
respectively.  Thus multiplying the average number of galaxies that
are found in the Photometry Catalog by $b$ results in the true
average.  This is not so apparent with other weighting methods, as
simply multiplying $\overline{W_{q}}$ by the inverse fraction of
galaxies detected with ground-based photometry assumes the missing
subset is representative of the entire ACS catalog.  Nonetheless,
given that the $N_{\rm ACS}$ and $N_{\rm phot}$ only differ at the
$5\%$ level, we expect the effects of such an assumption to be small.

\begin{figure}[t]
	\includegraphics[width = 0.5\textwidth]{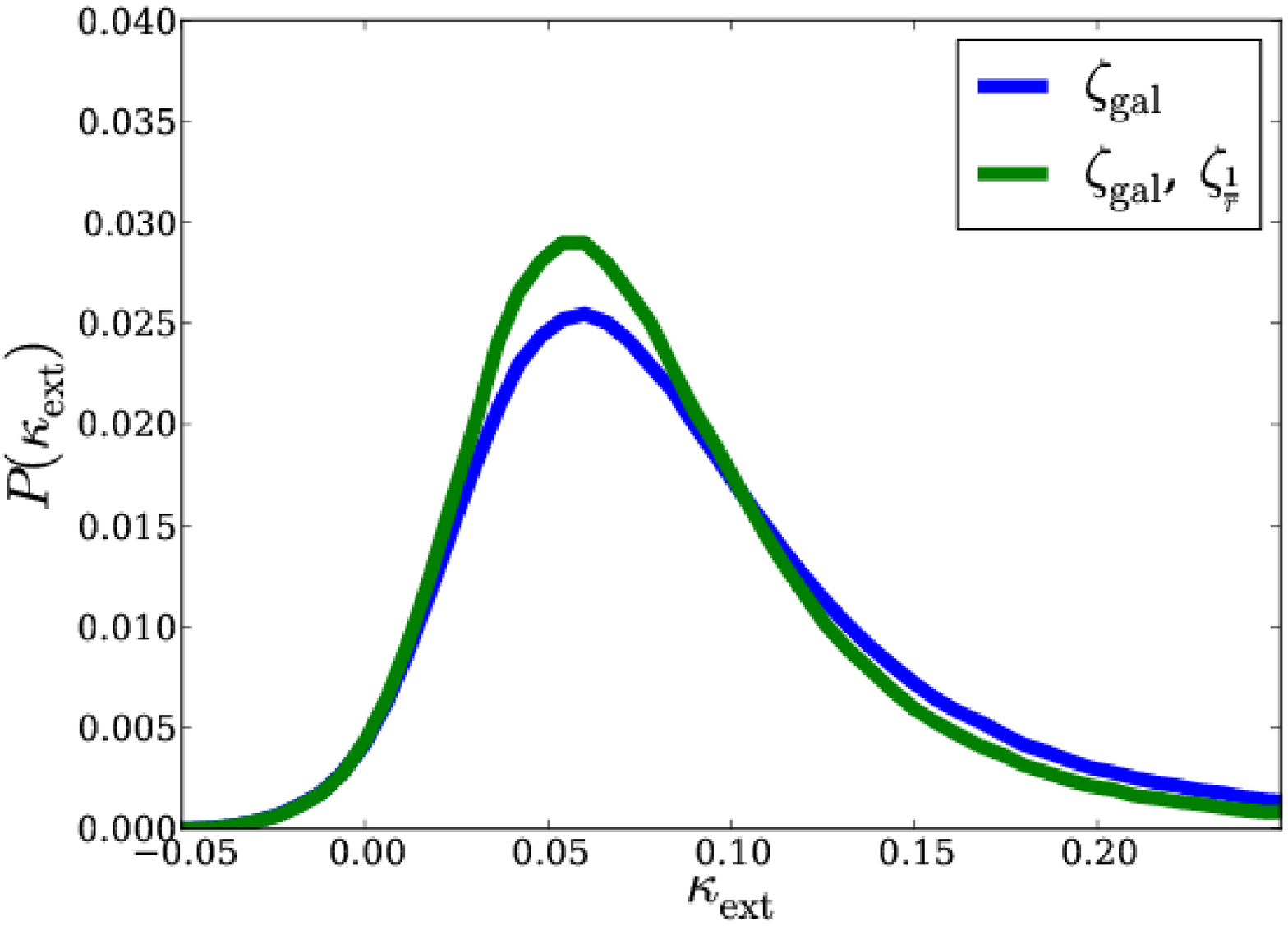}
	\caption{\label{fig:b1608pdf} $\kext$ PDFs for $\zetagal$ and $\zeta_{\frac{1}{r}}$ values as measured from \blens.}
\end{figure}
        
\begin{table*}[t]
	\begin{center}
        \caption {\label{btable} Statistics of \blens\ relative to COSMOS}
	\begin{tabular}{ccccccc}
		\hline
		$q$ & $\overline{W_{q}}$ & $W_{q}$ & $\zetaq$ & $\kmed$ & $\sigmak$ & $N_{\rm LOS}$\\
& (COSMOS) & (\blens) & & & &\\
		\hline
		$\rm gal$ & 41.6 & 83 & 1.997 & 0.0808 & 0.0562 & 2123167\\
		$\frac{1}{r}$ & 3.1799 & 6.2928 & 1.979 & 0.0741 & 0.0471 & 527734\\
%		$L$ & 0.7197 & 1.2111 & 1.683 & 0.0855 & 0.0546 & 134649\\
		$z$ & 15.6181 & 30.6207 & 1.961 & 0.0784 & 0.0534 & 328798\\
%		$\frac{L}{r}$ & 0.0551 & 0.0953 & 1.731 & 0.0730 & 0.0486 & 151441\\
		$\frac{z}{r}$ & 1.1947 & 2.3658 & 1.980 & 0.0695 & 0.0464 & 314931\\
		\hline
	\end{tabular}
        \end{center}
	Notes. Column 1 lists the statistic, Column 2 are the average weights
from lines of sight in COSMOS, Column 3 are the weights for \blens,
and Column 4 are the relative overdensities (the ratio of Column 3 to
Column 2).  Columns 5 and 6 are, respectively, $\kext$ and $\sigmak$
values found with all Millennium Simulation LOS that satisfy Equation
(\ref{eq:wcon}) for both $\zetagal$ and $\zetaq$ conditions imposed.
The number of LOS found from the Millennium data is also provided in
Column 7 to indicate the reliability of the data.
\end{table*}

To determine $W_{q}$ for \blens, we identify objects in the field of
\blens\ using SExtractor \citep{BertinArnouts96}.  
To ensure a fair comparison with COSMOS, we use only a single
\textit{HST} orbit of imaging (so that the COSMOS and \blens\ images
have similar depths), and follow the reduction steps outlined in
\citet{2007ApJS..172..219L}, with the exception of masking out 
asteroid trails, oversaturated stars, etc.~since these are not present
within $45''$ of \blens.  Once a catalog of all objects with $r <
45''$ and $18.5 < m_{\rm F814} < 24.5$ has been created, the objects
are coordinate-matched with a deeper catalog, constructed from the
full 11 \textit{HST} orbits of imaging.  This second catalog serves
two purposes: (1) to exclude fake detections from the single orbit
SExtractor data set, and (2) to produce more reliable redshifts and
stellar masses necessary to find $\zetaq$ for the detected objects in
the single-orbit catalog.  A limited number of objects ($\sim 15$)
have spectroscopic redshifts, or Gunn g, r, and i bands from
ground-based imaging.  Most, however, have just F814W and F606W
photometries that make it difficult to obtain accurate stellar masses
and photometric redshifts.

	\subsection{Finding $\zetaq$ for B1608}
	\label{ssec:bzetaq}

We impose $z < 1.4$ for COSMOS and \blens.  We find that the observed
average number of galaxies in COSMOS is $\overline{N_{\rm gal}} =
41.6$ while \blens\ has $83$ galaxies within $45''$, giving $\zetagal
= 1.99$.  This is close to $\zetagal = 2.18$ as found by \citet{FKW11}
by comparing the galaxy counts to those in pure-parallel fields.  The
slight difference in $\zetagal$ could be due to (1) our imposed
redshift restriction $z<z_{\rm source}$ that was not applied by
\citet{FKW11}, and (2) the COSMOS field being slightly overdense.
However, it is not clear how much COSMOS is overdense when the
redshift condition is imposed.  For simplicity, we neglect this
correction, though in the future this will need to be measured before
applying it to time-delay systems for cosmological inferences.  In a
method analogous to Section~\ref{sec:weights} we compute $\zetaq$ for
each characteristic, the results of which can be seen in
Table~\ref{btable} for characteristics that are computed with a higher
degree of accuracy (e.g., $1/r$). Unfortunately the present data are
not sufficient to estimate reliable stellar masses, luminosities, or
accurate photometric redshifts.  Deeper optical and NIR imaging of the
field are necessary to obtain more accurate redshifts (see
Table~\ref{zphotsigs}), luminosities, and stellar masses for computing
$\zetaq$.

Next we select from the Millennium Simulation lines of sight with the
new $\zetaq$ values to find $\kext$.  In keeping with
Section~\ref{sec:weights} we impose Equation (\ref{eq:wcon}) for both
$\zetagal$ and $\zetaq$ and find $\kext$ and $\sigmak$ for the
resulting distribution.  These, along with the number of lines of
sight, are given in the last three columns of Table \ref{btable}.
Distributions with $\zetagal \sim\zetaq$ are closely correlated and
therefore have large $N_{\rm LOS}$.  In Figure~\ref{fig:b1608pdf}, we
show the $\kext$ PDF for \blens\ with (1) only $\zeta_{\rm gal}$
imposed, and (2) both $\zeta_{\rm gal}$ and $\zeta_{1/r}$ conditions
imposed.  The additional $\zeta_{1/r}$ condition sharpens the PDF,
leading to a decrease in $\sigmak$ from 0.056 by $\sim 0.01$.  In this
specific case of \blens, the new $\kext$ PDF does not decrease the
uncertainty on the final time-delay distance measurement appreciably
since the stellar kinematics of the lens galaxy provides substantial
constraints on $\kext$ already, similar to the level that is achieved
with the multiple $\zetaq$ conditions.  Nonetheless, the PDF of the
time-delay distance is shifted to lower values by $1-2\%$ due to the
lower $\kmed$. We thus conclude that even though these effects are
smaller than current uncertainties for a single lens, they will become
important for reaching the ultimate goal of 1\% precision and
accuracy.

To generalize, without the velocity dispersion as a constraint, the new
$\kext$ would have decreased the uncertainty on the resulting $H_0$
from \blens\ for various uniform cosmological priors by $\sim 1\,
\rm{km\,s^{-1}\,Mpc^{-1}}$ from $\sim4 \,\rm{km\,s^{-1}\,Mpc^{-1}}$. 
Therefore, for lens systems in which the lens velocity dispersion is
difficult to obtain (due to, e.g., bright lensed images that are near
the lens galaxies), or for very large samples for which stellar
velocity dispersions might not be practical, our techniques for
tightening $\kext$ are especially valuable since the reduction in the
uncertainty on $\kext$ would then translate directly to that on the
time-delay distance (e.g., $0.01$ in $\kext$ is approximately 1\% on
the time-delay distance).

\section{SUMMARY AND CONCLUSIONS}
\label{sec:conclusion}

With the goal of finding ways to measure the effects of the
distribution of mass along the line of sight to gravitational lensing
time delays, we have performed a comprehensive analysis of simulated
lines of sight catalogs. These lines of sight catalogs are based on
the Millennium Simulation, used to compute the external convergence
$\kext$ via ray-tracing, as well as on semianalytic models of galaxy
formation, used to assign observable properties to halos along the
line of sight. 

Our main results can be summarized as follows

\begin{enumerate}

\item{The observed relative abundance of galaxies within a given aperture $\zetagal$ provides an estimate of $\kext$ that is accurate to a few percent, depending on the actual under/overdensity of the observed line of sight. This is consistent with previous work \citep{SuyuEtal10}.}
\item{Adding information from other observables like stellar mass, luminosity, redshift, position of the galaxies in the vicinity of the main deflector, reduces significantly the uncertainty in $\kext$. The most significant drop in uncertainty is obtained by weighting each galaxy with the inverse of the projected distance to the main deflector, followed by powers of the stellar mass. With this kind of information the uncertainty on time-delay distance arising from $\kext$ can be reduced to $\sim4$\% from $\sim 6$\% using only galaxy counts for a very overdense line of sight and to $\sim3$\% for typical lines of sight.}
\item{Even though spectroscopic redshifts are valuable, especially for the galaxies most closely associated with the main deflector, photometry in three optical bands (e.g., gri) and the near infrared (K) are sufficient for obtaining photometric redshifts such that the median and width of the $\kext$ change by $<0.007$, i.e., $<0.7\%$ on the time-delay distance.}
\item{As a practical illustration, we apply this method to the field of \blens\ and show that some gain can be made even with existing data. Better multiband photometry is needed to fully realize the gains promised by our method.}
\end{enumerate}

From these results, we conclude that with sufficient imaging and
spectroscopy data the effects of the mass distribution along the line
of sight on gravitational time-delay distances can be accounted for
and the associated uncertainties reduced for all lines of sight.
These improvements -- in combination with recent advances in the
derivation of gravitational time delays \citep{TewesEtal12} and in the
modeling of the mass distribution of the main deflector and objects in
close proximity to it \citep{SuyuEtal12} -- bring us closer to the
goal of 1\% precision in cosmological distances, necessary to address
fundamental issues such as the nature of dark energy
\citep{SuyuEtal12a}.  

In the next decade, upcoming surveys are expected to deliver thousands
of gravitationally lensed quasars \citep{OguriMarshall10}, a number
more than sufficient to meet the 1\% goal provided effort is made to
keep systematic uncertainties under control.  This will have to
include theoretical uncertainties related to the choice of numerical
simulations and associated semianalytic models. Our choice of using
overdensities with respect to random fields, as opposed to absolute
densities, minimizes the impact of the choice of this specific model.
However, as the number of lenses with measured time delays increases
thus reducing the observational errors toward the 1\% level, it will
be important to repeat and extend this study with independent
cosmological simulations and galaxy formation models. This is left for
future work.

From an observational point of view, the future abundance of targets
will change dramatically the situation with respect to the present
time when the precision of time-delay cosmology is limited by the
number of known strongly lensed quasars, and allow us to choose the
targets that give more cosmological information at fixed observational
resources. This work suggests that focusing follow-up efforts on
specific lines of sight -- those that are not too overdense with
respect to the average of the universe -- should result in substantial
gains in efficiency.

\acknowledgments We would like to thank Dan Coe for his help with BPZ,
and the anonymous referee for his/her constructive comments that
greatly improved the presentation of this work.  Z.S.G., S.H.S.~and
T.T.~gratefully acknowledge support from the Packard Foundation in the
form of a Packard Research Fellowship to T.T.  and from the National
Science Foundation grant AST-0642621. S.H. and R.D.B.~acknowledge
support by the National Science Foundation (NSF) grant number
AST-0807458. T.E.C. is supported by an STFC
studentship. P.J.M.~acknowledges support from the Royal Society in the
form of a research fellowship.  C.D.F.~acknowledges support from
NSF-AST-0909119. L.V.E.K. is supported in part by a NWO-VIDI program
subsidy (639.042.505). Support from HST program GO-12889 is gratefully
acknowledged. This work is based in part on observations made with the
NASA/ESA Hubble Space Telescope, obtained at the Space Telescope
Science Institute, which is operated by the Association of
Universities for Research in Astronomy, Inc., under NASA contract NAS
5-26555. These observations are associated with program \#GO-10158 and
\#GO-9822.  The Millennium Simulation databases used in this paper and
the web application providing online access to them were constructed
as part of the activities of the German Astrophysical Virtual
Observatory.

\end{document}